\documentclass[11pt]{article}
\usepackage[margin=1in]{geometry}
\usepackage[utf8]{inputenc}
\usepackage[T1]{fontenc}
\usepackage{graphicx,psfrag,epsf} 
\usepackage{listings}
\usepackage{enumitem}
\usepackage{multirow}
\usepackage{wrapfig}
\usepackage{lscape}
\usepackage{rotating}
\usepackage{epstopdf}
\usepackage{pdflscape}
\usepackage{moresize}
\usepackage[colorlinks=true, allcolors=blue]{hyperref}
\usepackage{cite}
\usepackage{amsmath}
\usepackage{algorithm}
\usepackage{algpseudocode}
\usepackage{array}
\usepackage{changepage}
\usepackage[english]{babel}
\usepackage{amsthm}
\usepackage{placeins}
\usepackage[autostyle=true]{csquotes}
\usepackage{multirow}
\usepackage{multicol}
\usepackage{mathrsfs}
\usepackage{stfloats}

\title{\bfseries Skill vs. Chance Quantification and Relative Ranking \\ for Popular Card \& Board Games}

\author{
\begin{tabular}{c}
Tathagata Banerjee \\
\small Indian Institute of Technology, Kanpur, India \\
\small \texttt{tathagatajoy5959@gmail.com}
\end{tabular}
\and
\begin{tabular}{c}
Anushka De \\
\small Indian Statistical Institute, Kolkata, India \\
\small \texttt{anushka.isical@gmail.com}
\end{tabular}
\and
\begin{tabular}{c}
Subhamoy Maitra \\
\small Indian Statistical Institute, Kolkata, India \\
\small \texttt{maitra.subhamoy@gmail.com}
\end{tabular}
\and
\begin{tabular}{c}
Diganta Mukherjee \\
\small Indian Statistical Institute, Kolkata, India \\
\small \texttt{digantam@hotmail.com}
\end{tabular}
}

\date{} 

\begin{document}
\maketitle
\begin{abstract}
This paper presents a data-driven statistical framework to quantify the role of skill in games, addressing the long-standing question of whether success in a game is predominantly driven by skill or chance. We analyze player-level data from four popular games—Chess, Rummy, Ludo, and Teen Patti—using empirical win statistics across varying levels of experience. By modeling win rate as a function of experience through a regression framework and employing empirical bootstrap resampling, we estimate the degree to which outcomes improve with repeated play. To summarize these dynamics, we propose a flexible skill score that emphasizes learning over initial performance, aligning with practical and regulatory interpretations of skill. Our results reveal a clear ranking, with Chess showing the highest skill component and Teen Patti the lowest, while Rummy and Ludo fall in between. The proposed framework is transparent, reproducible, and adaptable to other game formats and outcome metrics, offering potential applications in legal classification, game design, and player performance analysis.
\end{abstract}
\vspace{1em}

\section{Introduction}
Games, whether played online or offline, serve not only as entertainment but also as exercises in strategy, engaging players in decision-making, learning, and social interaction. The rapid growth of the online gaming industry driven by the ubiquity of smartphones and internet access has transformed games into a multi-billion-dollar business ecosystem. However, this expansion has also raised legal, economic, and ethical concerns, particularly when monetary rewards are involved. Ethical considerations in the games industry are gaining prominence, with ongoing discourse emphasizing the need for responsible design, player well-being, and inclusive development practices \cite{ethics2023roundtable}.

At the heart of legal and regulatory debates is the classification of games as either skill-based or chance-based, a distinction with far-reaching policy and financial implications. Game mechanics such as levels and high scores have been shown to significantly influence user engagement and persistence, highlighting the motivational power of structured progression systems within gamified experiences \cite{siriaraya2023motivational}. Yet, the boundary between skill and chance is often blurred, as most games incorporate elements of both. Thus, the critical question is not whether a game is purely one or the other, but rather: how much skill is involved in determining outcomes? Quantifying this continuum with empirical rigor is essential for informed debate, legal clarity, and greater industry transparency.

Several recent works have addressed this challenge using diverse approaches. Goodman et al. (2024) propose a simulation-driven framework to separate skill and randomness in tabletop games, emphasizing how tournament seeding impacts measurable outcomes \cite{Goodman_2004}. Duersch et al. (2020) provide an econometric variance decomposition model to quantify the relative importance of skill and chance, especially for regulatory classification of games \cite{DUERSCH_Lambrecht}. Borm and van der Genugten (2001) introduce a relative measure of skill for games with chance elements by comparing expected and observed success frequencies across player types \cite{Borm2001OnAR}. Getty et al. (2018) present a method to quantify the legal implications of randomness in fantasy sports using information-theoretic concepts \cite{Getty18Fantasy}. Levitt and Miles (2014) analyze extensive poker data to statistically demonstrate the presence of skill over repeated play \cite{Levitt}. Orkin (2021) explores the philosophical and mathematical boundaries between skill and chance, highlighting the role of statistical repeatability as an indicator of skill \cite{Game_Orkin}.
Several additional works have explored the complex boundary between skill and chance in games through various empirical and theoretical lenses. Agt et al. (2021) present a student-led empirical analysis comparing different games on a skill-chance spectrum using win/loss data and structural game features \cite{Agt} . Erdmann (2011) provides a theoretical framework for characterizing games based on the degree of player influence versus randomness, emphasizing philosophical and probabilistic perspectives \cite{Erdmann2011TheCO}. Hannum and Cabot (2009) argue for the legal classification of poker as a game of skill, synthesizing statistical evidence with legal criteria \cite{Hannum}. More recently, Stafford and Vaci (2022) discuss how digital games can be leveraged to study skill acquisition over time, advocating for fine-grained behavioral data in psychological research \cite{vaci}. These works underscore the diversity of approaches to measuring skill and support the need for a scalable, data-driven framework like the one proposed in this paper.

Building on this literature, our paper presents a statistically grounded, data-driven framework to quantify the role of skill in popular card and board games using real-world player-level data. Specifically, we analyze four games: \textit{Chess}, \textit{Rummy}, \textit{Ludo}, and \textit{Teen Patti} , to determine the extent to which experience and prior performance influence success. We define `skill' as a player's ability to consistently improve outcomes through learning, adaptation, and strategy, in contrast to outcomes primarily driven by randomness or fixed initial conditions.

Our contribution is threefold. Firstly, we propose a regression-based skill quantification model that uses transformed win rates and player experience as key inputs, accounting for both innate ability and learning. Secondly, we introduce an empirical bootstrap resampling procedure to assess the robustness and uncertainty of skill estimates across games, avoiding reliance on strong parametric assumptions. Finally, we construct a flexible skill score metric that allows for subjective weightings and cut-offs, enabling comparative ranking of games while maintaining transparency about the assumptions involved.

The framework is applied to large-scale proprietary datasets (Rummy, Ludo, Teen Patti) and public datasets (Chess) to derive skill scores for each game. Our results show that while Chess predictably scores highest on the skill spectrum, games like Rummy and Ludo also demonstrate significant skill components—contrary to popular perception—whereas Teen Patti shows relatively weak evidence of strategic learning.

The remainder of this paper is structured as follows. Section II presents the algorithmic framework in a step-by-step format to enhance transparency and reproducibility. Section III outlines the empirical methodology, including the model specification, estimation strategy, and resampling techniques used to derive robust skill estimates. In Sections IV through VII, we present game-wise statistical analyses of Chess, Rummy, Ludo, and Teen Patti, respectively—detailing both regression and bootstrap results. Section VIII introduces our flexible scoring methodology to convert statistical findings into interpretable skill scores and provides comparative rankings. Finally, Section IX discusses key insights, limitations, and potential extensions, including the applicability of the framework to alternative measures of game success.

\subsection{The Game Structures Considered} 
The quantification of skill vs. chance in a game has several inputs. We arrange these in a logical manner as below.

\begin{enumerate}

\item {\sf Randomisation only at the beginning:} we start with the unlikely example of Chess, which is considered to be a purely skill based game. The only role of chance is playing with Black or White pieces, which is statistically shown to have some implication in terms of winning ceteris paribas (a Latin phrase that means ``all other things being equal" or ``other things held constant").

A second, and more common example is that of Rummy, where the role of chance is in the initial permutation of the cards. Then a partition of which results in the two hands with the players and the sequence in the deck. Here, it is shown that there is significant skill (experience) premium in the Rummy game. 

\item {\sf There is randomness in game play also:} Ludo is a prime example where the players start with identical initial positions but nature's move (roll of the dice) dictates the choices available to them at each turn. Again, we show that substantial skill premium (benefit from the use of smart strategy) exists in Ludo.

\item {\sf A second aspect is observability:} whether the players can see opponent's position fully (as in Chess or Ludo) or not (in Rummy). This results in mental probability/belief updations that are refined through the course of game play. this is another aspect of Rummy that requires substantial skill. 

\item Thus one may think of variations on the existing games which would highlight one or the other aspects (randomisation, observability) of a game more. Consider the Ludo Ninja variation (\url{https://www.zupee.com/ludo/ludo-ninja/}) which decides the rolls of the die for each turn at one go. At the very beginning, a sequence of numbers in $\{1, 2, 3, 4, 5, 6\}$ are given to each player. So there is no randomisation after the start. The sequence for the opponent is unknown to a player. So there is also incomplete observability. In this sense, Ludo Ninja is closer to Rummy. This variation of Ludo also benefits from the use of smart strategies.

\item One may also think of a further variation where both the lists will be observable to the players, so that there is no lack of observability and then the game becomes one of pure strategy like Chess. Experimentation is required to understand the value of skill in such versions.
\end{enumerate}

Thus, how do we proceed towards this quantification? We adopt an approach similar, but not identical, to \cite{Getty18Fantasy} in the context of fantasy games or that of \cite{Levitt} for Poker. Our approach will be more statistically oriented, through a regression type analysis. 
For this purpose it is imperative to create a benchmark for Chess, which is considered closest to a pure skill game. We will use statistical methodology similar to that used for Rummy in the context of the game of Chess where skill will be proxied by ELO rating and age (in a non-monotonic manner, see \cite{Kalwij}). The advantages of this approach is {\bf threefold}. First, the change dynamics of performance is captured in terms of only a few parameters (see equation (\ref{diffeqn}) below) like experience as proxy for learning, previous performance as proxy for innate skill and current performance to model saturation (fatigue). Secondly, we use a linearised version of this for empirical purposes that enables the use of the easily understood linear regression methodology (see equation (\ref{regeqn}) below) that allows the study of partial effects easily. Thirdly, our skill quantification formulation allows sufficient flexibility in choosing the sensitivity with respect to each of these parameters in terms of scaling and critical cut-offs (see eqn (\ref{scalingeqn}) below) as well as their relative importance (see equation (\ref{wtgeqn}) below). This flexibility results in a distribution of skill scores rather than a single number, as evidenced in table \ref{score_comparison} and figures \ref{fig:Chess_Scores_Transformations}, \ref{fig:Rummy2_Scores_Transformations}, \ref{fig:Ludo_Scores_Transformations} and \ref{fig:TeenPatti_Scores_Transformations}. Is this unsatisfactory? We claim that this is more useful than a single number as opinions vary widely in terms of relative importance and scaling of these parameters in the skill measurement literature. Our framework provides, to the best of our knowledge, the first objective data driven framework to pursue this debate in a scientific manner.   

\ \\
{\bf Data:}
For Rummy, Ludo and Teen Patti, we use proprietary data from OSG companies (in particular {\it Games 24X7}, \url{https://www.games24x7.com/}). For benchmarking purpose we also do extensive simulation studies. For Chess, we use publicly available game level outcome data from FIDE official website (\url{https://www.fide.com/}).
\section{Algorithmic Approach}

To enhance transparency and reproducibility, we outline below the step-by-step procedure followed in our analysis. This algorithmic presentation is designed to be easily understood, even by readers without a technical background.

\subsection*{Step 1: Data Preparation}
\begin{itemize}
    \item Collect player-level data for each game, including the number of games played and corresponding win rates.
    \item Organize data into intervals based on the number of games played (e.g., 1--10 games, 11--20 games, etc.).
\end{itemize}

\subsection*{Step 2: Experience Grouping}
\begin{itemize}
 \item  Divide players into experience quantiles according to the number of games played.
\item For each quantile, compute the outcome (e.g., average win rate) to track how performance evolves with experience.
\item Samples are required to be created at a user level across different experience partitions and/or within partitions to create enough variety for the modeling purposes
\item One manifestation of this could be to consider the experience (i.e., the no. of games) into two halves and try to infer the second half outcome as a function of the first one. This can be performed across different experience levels to ensure the variety (e.g., players playing 100 games being partitioned into 1st 50 and 2nd 50, players playing 500 games being partitioned into 1st 250 and 2nd 250, and so on).

\end{itemize}
\subsection*{Step 3: Regression Modeling}
\begin{itemize}
    \item Fit a regression model where the dependent variable is the win rate and the independent variable is the experience level. \textit{(The model is at a user level. We are using the groupings in the third point of step 2. Partitions refer to the two halves based on which player performance and experience are considered.
)}
    \item Estimate the slope to quantify the skill-growth trend - how much a player’s win rate improves with additional experience.
\end{itemize}
\subsection*{Step 4: Bootstrap Resampling}
\textit{The buckets refer to the groupings in the third point of step 2. Resampling is done over the set of all users as we have performance and experience data of two halves for all users.
}
\begin{itemize}
    \item Apply the empirical bootstrap technique within each experience bucket to estimate the uncertainty around the average win rate.
    \item For each resample, recalculate the average win rate to generate a distribution of estimates.
\end{itemize}

\subsection*{Step 5: Confidence Interval Construction}
Confidence intervals are needed to provide the estimate of probability perspective (uncertainty, fluctuations etc.) of a model.
\begin{itemize}
    \item Use bootstrap distributions of regression slopes to construct confidence intervals. \textit{ (Construction of confidence intervals using bootstrap is preferred since it doesn’t require any parametric assumptions.)}
    \item These intervals quantify the statistical uncertainty associated with the estimated skill-growth rate.
\end{itemize}

\subsection*{Step 6: Normalize Skill Estimates}
For comparability across games and avoid possible scaling issues, a data dependent normalisation of the estimates is necessary. 
\begin{itemize}
    \item For each game, normalize the bootstrap slope estimates by dividing them by the average win rate in the first experience half.
    \item This yields a \textit{relative skill slope} that accounts for baseline win rate differences across games.
\end{itemize}

\subsection*{Step 7: Compute Summary Score}
\begin{itemize}
    \item The median of the outcome measure (e.g., win rate) can be considered as a summary, as it is more robust to outliers than the conventional average value.
\textit{The model was run a large number of times on resampled datasets, and median here refers to the median of the summary statistics from the repeated executions of the model.}
    \item This median represents the \textbf{Final Skill Score}—a single interpretable value capturing the game’s skill component.
    \item Compute the median of the relative slope distribution for each factor for each game.

\end{itemize}

\subsection*{Step 8: Cross-Game Comparison}
\begin{itemize}
    \item Use the Final Skill Scores to rank or compare games in terms of how much skill contributes to success.
    \item Higher scores suggest greater influence of skill, while lower scores may imply chance, volatility, or short learning curves.
\end{itemize}
\section{Framework for Empirical Analysis}
\subsection{Data Format}
To begin with consider having a games outcome data, measured in terms of win/loss for various games for multiple players. 
For an empirical investigation of the robustness of our findings, we need a tournament setting where a large pool of players play against each other. These results can be used to empirically define the variability (or otherwise) of our results using resampling techniques.
\subsection{Our Luck vs. Law measurement formulation}
\label{LL}
After suitably identifying the data, a suitable statistical model is needed to determine whether the game is of skill or chance. 

The model we propose is as follows:
\begin{itemize}
    \item Assume the following two factors are important for a favourable outcome:
    \begin{itemize}
    \item Innate skill: $h$, unobservable
    \item Learning $l_t = l(e_t)$, where $e_t$ is experience, measured by the number of games previously played and so it is observable. 
    \end{itemize}
    Learning rate can be different between players (efficiency). However, one needs to consider the fatigue too.
    \begin{itemize}
    \item Realised Skill $s_t = s(h, l_t):$ also unobservable
    \item Strategy for each player: $\sigma_1, \sigma_2$ etc.
    \item Outcome (win/loss) $w_t = w(\sigma_1 (s_{1t}), \sigma_2 (s_{2t}), \epsilon_t) = w(\sigma_1 (s(h_1, l_{1t})), \sigma_2 (s(h_2, l_{2t})), \epsilon_t)$. \\ Here $\epsilon_t$ is the environmental uncertainty (exogenous to the players' actions).
    \end{itemize}
    \item The choice of strategy depending on skill creates an additional perceived uncertainty in choice of $\sigma_{1 (2)}$ for player 2 (1). But this is actually part of skill.
\end{itemize}
\subsection{Estimation Strategy}
 As skill has unobservable inputs, a direct estimation of the above relationship is not possible from (experience, outcome) data. So we posit the following: \\
$dw_t = \frac{\partial w}{\partial h} + \frac{\partial w}{\partial l_t}$ and $dw_{t-1} = \frac{\partial w}{\partial h} + \frac{\partial w}{\partial l_{t-1}}.$ Differencing yields $dw_t - dw_{t-1} = \frac{\partial w}{\partial l_t} - \frac{\partial w}{\partial l_{t-1}}$.
Denote $\Delta e_t= e_t- e_{t-1}$.

As we are looking for an intuitive and easily computable benchmarking procedure, we stick to a linearized relationship. Hence, we posit the following regression equation to represent a linear approximation of the above Differential Equation: 
\begin{equation}
\label{diffeqn}
    w_t = \alpha +\beta_1 w_{t-1} + \beta_2 e_{t-1} + \beta_3 \Delta e_t + \text{error} 
\end{equation}
(here experience works as a proxy for learning and the outcome represents the chance).

Consider two time intervals (0, 1) and (1, 2). Denote by $m_i, \, i = 0, 1, 2$ experience (games previously played) at these time points. 
Calculate $e_1=(m_0 + m_1)/2$ and $e_2=(m_1 + m_2)/2 - m_1 = (m_2 - m_1)/2$. Denote the win percentage in intervals (0,1) and (1,2) as $w_1$ and $w_2$. 

\textbf{Practical Application: }
For empirical convenience, we will use the following transformed version of the regression equation, that spans the whole real line:
\begin{equation}
\label{regeqn}
    \Phi^{-1}(w_2) = \alpha_0 +\alpha_1 \Phi^{-1}(w_1) + \beta_1 e_1 + \beta_2 e_2 + \text{error}.
\end{equation}
, where $\Phi^{-1}(w)$ is the represents the inverse of the standard normal cumulative distribution function, also known as the quantile function.
\subsection{Assessing Robustness using Resampling}
We employed the empirical bootstrap technique to address the regression problem, enabling robust estimation of variability without relying on strong parametric assumptions. In this case, the empirical bootstrap is also called paired bootstrap. Consider given a paired sample: $\left(X_{1}, Y_{1}\right), \cdots, \left(X_{n}, Y_{n}\right)$, (where $X_i, Y_i$ represent some data points) we generate new sets of IID observations
$$
\left(X_{1}^{*}, Y_{1}^{*}\right), \cdots,\left(X_{n}^{*}, Y_{n}^{*}\right)
$$
such that for each $\ell$,
$$
P\left(X_{\ell}^{*}=X_{i}, Y_{\ell}^{*}=Y_{i}\right)=\frac{1}{n}, \quad \forall i=1, \cdots, n.
$$
Namely, we treat $\left(X_{i}, Y_{i}\right)$ as one object and sample with replacement $n$ times from these $n$ objects to form a new bootstrap sample. Thus, each time we generate a set of $n$ new observations from the original dataset.
Assuming the entire process is repeated $B$ times, we would obtain
$$
\left(X_{1}^{*(1)}, Y_{1}^{*(1)}\right), \cdots,\left(X_{n}^{*(1)}, Y_{n}^{*(1)}\right) \\
\left(X_{1}^{*(2)}, Y_{1}^{*(2)}\right), \cdots,$$ $$\left(X_{n}^{*(2)}, Y_{n}^{*(2)}\right) \left(X_{1}^{*(B)}, Y_{1}^{*(B)}\right), \cdots,\left(X_{n}^{*(B)}, Y_{n}^{*(B)}\right)$$
For each bootstrap sample, say $\left(X_{1}^{*(\ell)}, Y_{1}^{*(\ell)}\right), \cdots, \left(X_{n}^{*(\ell)}, Y_{n}^{*(\ell)}\right)$, we fit the linear regression, leading to a bootstrap estimate of the fitted coefficients $\widehat{\beta}_{0}^{*(\ell)}, \widehat{\beta}_{1}^{*(\ell)}$. Thus, the $B$ bootstrap samples leads to
$$
\left(\widehat{\beta}_{0}^{*(1)}, \widehat{\beta}_{1}^{*(1)}\right), \cdots,\left(\widehat{\beta}_{0}^{*(B)}, \widehat{\beta}_{1}^{*(B)}\right)
$$
$B$ sets of fitted coefficients. We then estimate the variance by

$\widehat{\operatorname{Var}}_{B}\left(\widehat{\beta}_{0}\right)  =\frac{1}{B} \sum_{\ell=1}^{B}\left(\widehat{\beta}_{0}^{*(\ell)}-\bar{\beta}_{0}^{*}\right)^{2} , \quad \bar{\beta}_{0}^{*}=\frac{1}{B} \sum_{\ell=1}^{B} \widehat{\beta}_{0}^{*(\ell)} $
,$\widehat{\operatorname{Var}}_{B}\left(\widehat{\beta}_{1}\right) =\frac{1}{B} \sum_{\ell=1}^{B}\left(\widehat{\beta}_{1}^{*(\ell)}-\bar{\beta}_{1}^{*}\right)^{2}, $
$\quad \bar{\beta}_{1}^{*}=\frac{1}{B} \sum_{\ell=1}^{B} \widehat{\beta}_{1}^{*(\ell)}$.

We can construct the confidence intervals using the variance estimate:
$$
\text { C.I. }\left(\beta_{0}\right)=\widehat{\beta}_{0} \pm z_{1-\alpha / 2} \cdot \sqrt{\widehat{\operatorname{Var}}_{B}\left(\widehat{\beta}_{0}\right)}, $$
$$ \text { C.I. }\left(\beta_{1}\right)=\widehat{\beta}_{1} \pm z_{1-\alpha / 2} \cdot \sqrt{\widehat{\operatorname{Var}}_{B}\left(\widehat{\beta}_{1}\right)}. $$
In summary, the application of the empirical bootstrap framework allows us to reliably estimate the variability in skill indicators across players with differing experience levels, without relying on strong distributional assumptions. This lays a statistically sound foundation for comparing skill dynamics across different games in the subsequent sections
\section{Analyzing Chess}
Chess is a two-player, turn-based strategy game played on an $8\times8$ board. It is a game of perfect information, meaning all elements of the game state are visible to both players at all times. The objective is to checkmate the opponent’s king. With no inherent randomness and a wide decision space, Chess is widely recognized as a skill-intensive game requiring deep planning and pattern recognition.

We present the analysis for various cases as follows.

\subsection{Classical}
For Classical format Chess, the leading FIDE tournament data is used. The downside of this being very small number of common players (the tournament only allows the top 14 players, see table \ref{tab:Chess_classical_data}). Thus results are not very reliable. In fact, none of the explanatory factors turn out to be significant as the p-value is greater than $5\%$ for all the covariates(see table \ref{tab:Chess_classical}). So we continue to a version with a larger pool of common players.

\begin{table}[h!]
\centering
\resizebox{\columnwidth}{!}{%
\begin{tabular}{|l|p{2cm}|p{2cm}|p{2cm}|p{2cm}|p{2cm}|p{2cm}|}
\hline
\textbf{Player} & \textbf{1st Year Win \% ($w_1$)} & \textbf{2nd Year Win \% ($w_2$)} & \textbf{$\phi^{-1}(w_1)$} & \textbf{$\phi^{-1}(w_2)$} & \textbf{1st Year Rating ($e_1$)} & \textbf{2nd Year Rating ($e_2$)} \\
\hline
Abdusattorov Nodirbek & 0.65 & 0.61 & 0.39 & 0.29 & 7.27 & 7.13 \\
Ding Liren            & 0.46 & 0.42 & -0.096 & -0.19 & 7.80 & 8.11 \\
Giri Anish            & 0.65 & 0.65 & 0.39 & 0.39 & 7.49 & 7.64 \\
Gukesh D              & 0.65 & 0.42 & 0.39 & -0.19 & 7.25 & 7.25 \\
Maghsoodloo Parham    & 0.35 & 0.54 & -0.39 & 0.096 & 7.40 & 7.19 \\
Praggnanandhaa R      & 0.58 & 0.46 & 0.19 & -0.096 & 7.43 & 6.84 \\
Van Foreest Jorden    & 0.35 & 0.46 & -0.39 & -0.096 & 6.82 & 6.81 \\
L'ami Erwin           & 0.65 & 0.50 & 0.39 & 0 & 6.27 & 6.27 \\
Roebers Eline         & 0.15 & 0.23 & -1.02 & -0.73 & 3.81 & 3.61 \\
Yilmaz Mustafa        & 0.39 & 0.69 & -0.29 & 0.502 & 6.65 & 6.09 \\
\hline
\end{tabular}
}
\caption{Player Statistics. There are only 10 common players combining Tata Masters and Challengers for Classical Chess, each have played 13 games in each year}
\label{tab:Chess_classical_data}
\end{table}

    \begin{table}[h!]
        \centering
        \resizebox{\columnwidth}{!}{%
        \begin{tabular}{|c|c|c|c|c|}
        \hline
        \textbf{Variable} & \textbf{Estimate} & \textbf{Std. Error} & \textbf{Test Statistic} & \textbf{p-value}\\
        \hline
    Intercept & -1.1106 & 0.8476 & -1.310 & 0.238  \\
    $\phi^{-1}$(First Year Win Proportion) & 0.1811 & 0.2912 & 0.622 & 0.557\\
    First Year Rating & 0.5363 & 0.4034 & 1.329 & 0.232\\
    Second Year Rating & -0.3797 & 0.3767 & -1.008 & 0.352\\   
     \hline
     \end{tabular}
     }
        \caption{Linear Regression Summary: Multiple $R^{2}$  = 0.483}
        \label{tab:Chess_classical}
    \end{table}

\subsection{Rapid, one year data}
In the version of Chess, there is a broader format that allows a larger number of players to play against each other. Now we first use only one year, a single edition of the tournament data. Since the number of games is $<$ 13 for some players (maybe they quit after some rounds), we consider both half game counts as predictors and then dropped the second half as it's not significant (see the result in table \ref{tab:Chess_rapid1}).
Now the explanatory factors return significant values (p values being less than 0.05) with ratings having a positive effect, but performance in the first half becomes negative. This could be due to the correlation between performance and rating. To understand this better, we move to a two-year analysis next.

\begin{table}[h!]
\centering
\resizebox{\columnwidth}{!}{%
\begin{tabular}{|c|c|c|c|c|}
\hline
\textbf{Variable} & \textbf{Estimate} & \textbf{Std. Error} & \textbf{Test Statistic} & \textbf{p-value}\\
\hline
Intercept & -3.08 & 0.47 & -6.51 & 0\\
$\phi^{-1}$(First Half Win Proportion) & -0.36 & 0.08 & -4.51 & 0\\
Rating & 0.16 & 0.021 & 7.52 & 0\\
First Half Games Count & 0.29 & 0.06 & 4.65 & 0 \\
\hline
\end{tabular}
}
\caption{Regression Summary: Multiple $R^{2}$ = 0.266. }
\label{tab:Chess_rapid1}
\end{table}

\subsection{Rapid, two year data}
In this extended version of the analysis using two years' data, we have a total of 96 players common in both editions of the tournament. Initial and final ratings denote the ratings in 2022 and 2023 respectively, and the response is $\phi^{-1}$ (win proportion in 23). The results in Table \ref{tab:Chess_rapid2} show that performance, initial rating, and second-year rating all have positive coefficients, which aligns with the expectation that both prior ratings and past performance significantly influence current outcomes (with p-values less than 0.05)

\begin{table}[h!]
\centering
\resizebox{\columnwidth}{!}{%
\begin{tabular}{|c|c|c|c|c|}
\hline
\textbf{Variable} & \textbf{Estimate} & \textbf{Std. Error} & \textbf{Test Statistic} & \textbf{p-value} \\
\hline
Intercept        & -1.09          & 0.14  & 7.99           & 0 \\
Initial Rating  & 0.34           & 0.06   & 5.68          & 0 \\
Final Rating    & 0.17         & 0.05  &  3.39 & 0.001 \\
$\phi^{-1}$(Initial Win Proportion)     & 0.12           & 0.09  & 1.41 & 0.163 \\
\hline
\end{tabular}
}
\caption{Regression Model Summary: R$^2$ = 0.611}
\label{tab:Chess_rapid2}
\end{table}

\subsection{Online Chess Data}
\subsubsection{Regression Results}
For this study, data has been collected from Lichess (\url{https://lichess.org/}), an online chess platform. Data of blitz games from March and April 2024 have been used, and only games played by players rated 2300+ on Lichess have been analyzed due to the availability of publicly accessible detailed data as opposed to players with lower ratings from whom public data is not available. Further, only the players who have played at least 5 games in each month have been considered for the study. The database consists of data for 5267 such players.

Since Lichess rating is updated after every game (unlike FIDE tournaments, where rating is not updated between tournament games), the experience gained in the months of March and April (considered to be the two halves) has been quantified in the following way.
 \begin{itemize}
     \item {\it First Half Experience} = $\frac{m_{0} + m_{1}}{2}$, where $m_{0}$ and $m_{1}$ are the ratings after the first and last matches in March respectively.
     \item {\it Second Half Experience} = $\frac{m_{2} + m_{3}}{2} - m_{1}$, where $m_{2}$ and $m_{3}$ are the ratings after the first and last matches in April respectively.
     \item Note that the players might have unrecorded matches between the calculation of $m_{1}$ and $m_{2}$, since the data only contains games where both players are rated 2300+. 
 \end{itemize}
It is to be noted that Lichess ratings and FIDE ratings are calculated slightly differently, and we perform origin-scale transformation on the experiences before using them for regression. Both the quantities used to quantify the experience in either half have been scaled to a range of 0 to 10. The regression summary is provided in table \ref{tab:chess_online_data}.

\begin{table}[h!]
\centering
\begin{tabular}{|c|c|c|c|c|}
\hline
Variable & Estimate & Std. Error & t-statistic & p-value \\
\hline
Intercept & -1.4001 & 0.0679 & -20.61 & 0 \\
$\phi^{-1}$(First Half Win \%)    & 0.1255  & 0.0138 & 9.12  & 0 \\
First Half Experience & 0.206  & 0.0102 & 20.23  & 0 \\
Second Half Experience & 0.1776  & 0.0151 & 11.75  & 0 \\
\hline
\end{tabular}
\caption{Regression Model Summary: R$^2$ = 0.574}
\label{tab:chess_online_data}
\end{table}
All three predictors—first-half win proportion, first-half experience, and second-half experience—have positive and statistically significant coefficients ($\text{p-value} < 0.05$). This suggests that both prior performance and accumulated experience contribute meaningfully to predicting current outcomes. The relatively high $R^2$ value (0.574) also indicates that the model explains a substantial portion of the variance in win rates among high-rated players, reinforcing the strong skill component in Chess.
\subsubsection{Bootstrap Results}
\begin{figure}[h!]
    \centering
    \includegraphics[width= 0.5\textwidth]{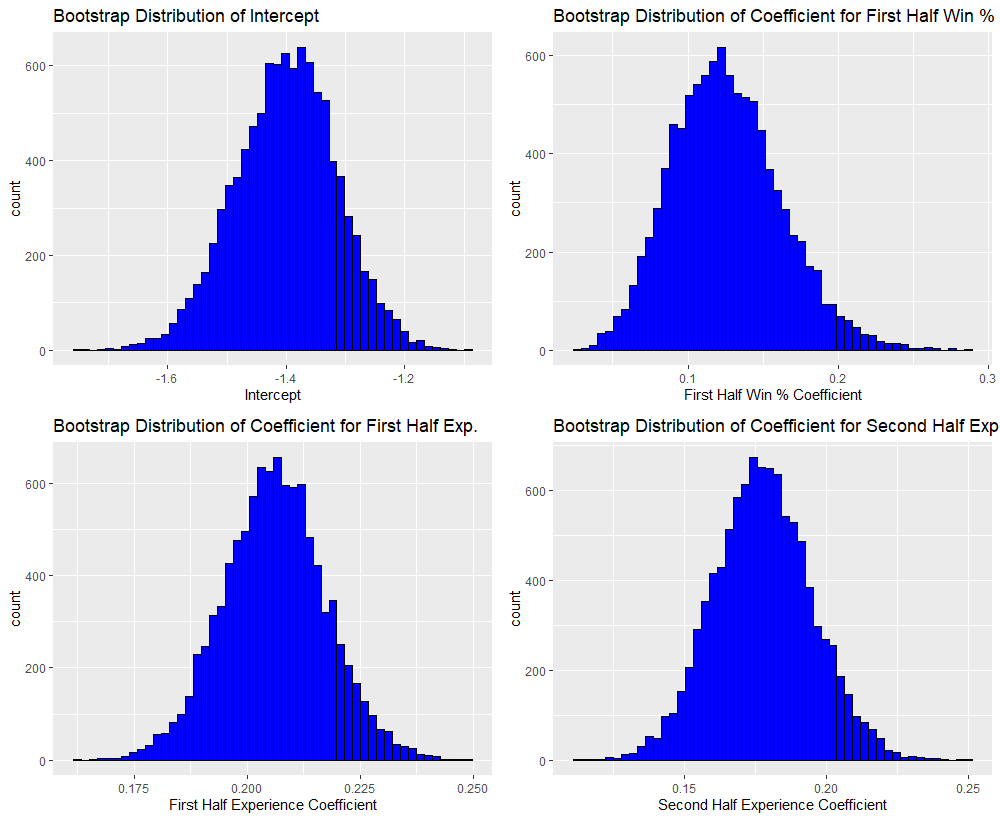} 
    \caption{Histogram of Bootstrap Estimates of Regression Coefficients}
    \label{fig:myplot}
\end{figure}

For performing resampling, 10000 bootstrap samples were drawn with replacement from the Lichess online database used for the study, each having size the same as the original dataset. The regression was performed on all of the resampled datasets. The bootstrap mean, variances and $95\%$ confidence intervals for the regression coefficients are provided in table \ref{tab:chess_online_data_boot} and the histograms are presented in the figure \ref{fig:myplot}.

\begin{table}[h!]
\centering
\resizebox{\columnwidth}{!}{%
\begin{tabular}{|c|c|c|c|c|}
\hline
Variable & Estimate & Bootstrap Mean & Bootstrap Variance & Bootstrap C.I\\
\hline
Intercept & -1.4001 & -1.4008 & 0.0034 & (-1.566, -1.217)\\
$\phi^{-1}$(First Half Win \%) & 0.1255  & 0.1255 & 0.0013 & (0.0572, 0.1795)\\
First Half Experience & 0.206 & 0.2061 & 0.0001 & (0.193, 0.2183)\\
Second Half Experience & 0.1776  & 0.1778 & 0.0003 & (0.1324, 0.2009)\\
\hline
\end{tabular}
}
\caption{Bootstrap regression summary for 10000 bootstrap estimates}
\label{tab:chess_online_data_boot}
\end{table}

The bootstrap distribution for Chess displays a consistently positive slope across resamples, with a narrow spread around the median. This confirms that increased experience is strongly associated with improved performance, supporting the high skill component in Chess.
\section{Analyzing Two Player Rummy}
Two Player Rummy is a card game played by 2 players using one or more standard decks of 52 cards (sometimes with jokers). The objective is to form valid combinations of cards called sets (same rank, different suits) or sequences (consecutive cards of the same suit). Each player is dealt a fixed number of cards, and players take turns drawing and discarding to improve their hands. The game incorporates both strategic planning and memory, as players must decide which cards to retain or discard based on observed play. Elements of chance exist through the card draw, but long-term success is strongly influenced by pattern recognition, probability estimation, and timing.

For the analysis of this card game we obtain actual data from online games. We analyze the data for 20,000 players, and the results are in table \ref{tab:Rummy}. 
The results show strong impact of experience in the face of performance. Current experience also has some impact (t-statistic being more than 1).
\begin{table}[h!]
    \centering
    \resizebox{\columnwidth}{!}{%
    \begin{tabular}{|c|c|c|c|c|}
    \hline
         Variable&   Estimate&  Std. Error&  t statistic& p-value\\
         \hline
         Intercept&  5.371e-02&  1.938e-03&  27.716& $<$2e-16 ***\\
         $\phi^{-1}(w_1)$&   1.092e-01&  6.936e-03&  15.738& $<$2e-16 ***\\
         First Half Experience 
&  8.109e-07&  3.773e-07&  2.149& 0.0316 *\\
         Second Half Experience &   7.101e-07&  6.669e-07&   1.065& 0.2870\\
         \hline
    \end{tabular}
    }
    \caption{Regression Summary for Rummy, multiple $R^2$:  0.013, based on 20,000 sample users}
    \label{tab:Rummy}
\end{table}
\FloatBarrier
 All variables except second-half experience are statistically significant at the 5\% level. The first-half win proportion has a particularly strong and significant positive impact, indicating the importance of innate or early-stage skill. First-half experience also shows a positive effect, suggesting that learning through play enhances performance. The model’s modest $R^2$ value (0.013) reflects the inherent randomness in card draws and other unobserved factors, but the significance of predictors still supports a meaningful role for skill in Rummy.

\subsection{Bootstrap results}
\begin{figure}[h!]
    \centering
\includegraphics[width=0.5\textwidth]{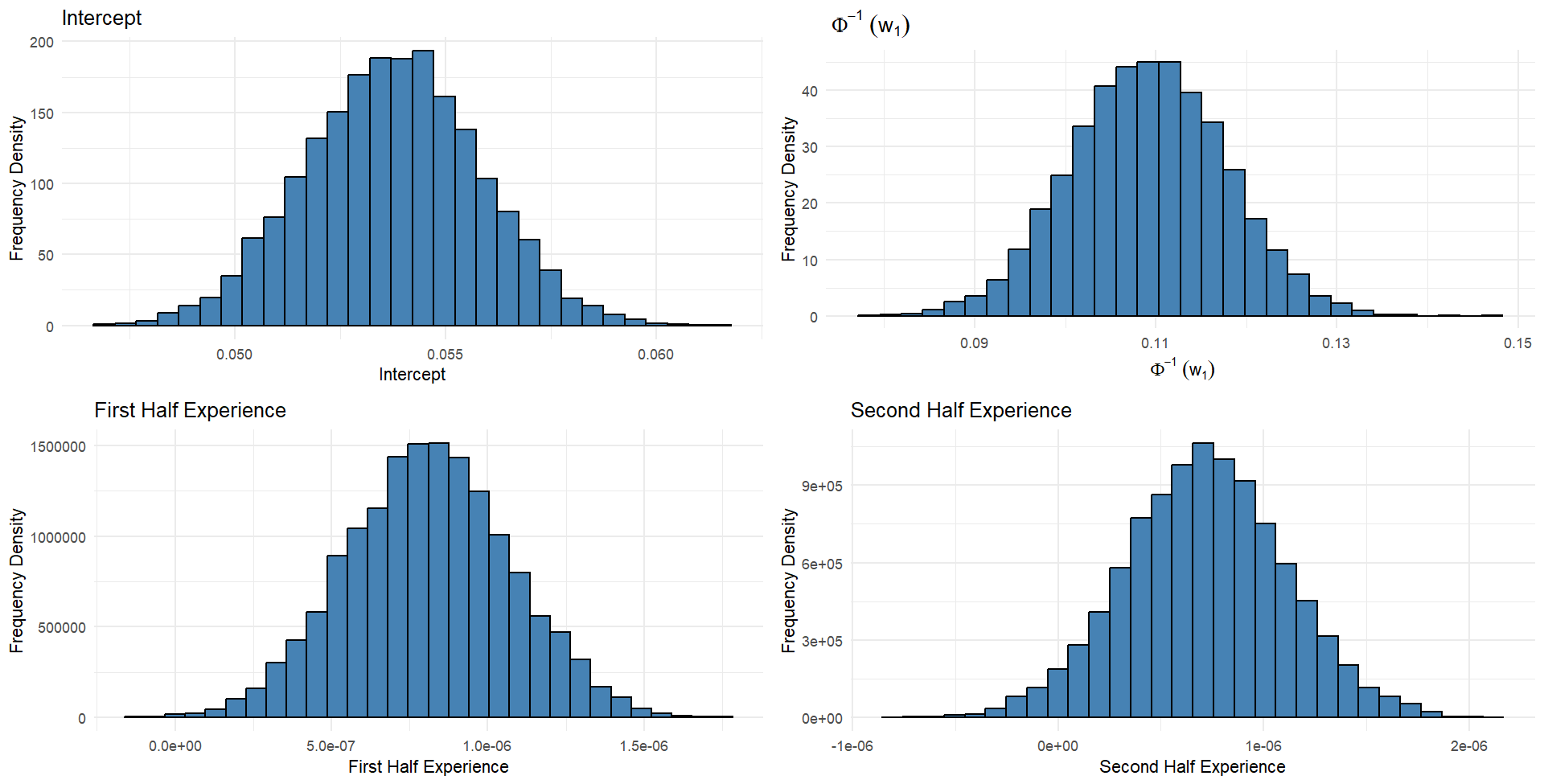}
    \caption{Bootstrapping Results for 2-Player Rummy Regression Model}
    \label{fig:Rummy2_Bt_Coef}
\end{figure}

The resampling results for Rummy (table \ref{tab:Rummy2 Bootstrap Regression Summary} and figure \ref{fig:Rummy2_Bt_Coef}) show very consistent patterns (very small standard errors and tight support), further supporting the robustness of our skill findings.
\begin{table}[h!]
    \centering
    \resizebox{\columnwidth}{!}{%
    \begin{tabular}{|c|c|c|c|c|}
    \hline
         Variable&   Estimate&  Bootstrap mean&  Bootstrap variance & Bootstrap $95\%$ CI\\
         \hline
         Intercept&  0.05371&  0.05371&  4.4496e-06& $(0.0496, 0.0577)$\\
         $\phi^{-1}(w_1)$&  0.1091&  0.10915&   7.2609e-05 & $(0.0924,0.125)$\\
         First Half Experience 
&  8.109e-07& 8.082e-07&  $~0$& - \\
         Second Half Experience &   7.101e-07&  7.163e-07&   $~0$& -\\
         \hline
    \end{tabular}
    }
    \caption{Bootstrap Regression Summary for $B=10000$ bootstrap iterations for $n=20000$ observations}
    \label{tab:Rummy2 Bootstrap Regression Summary}
\end{table}
The bootstrap distribution for Rummy shows a consistently positive slope, with most samples indicating a clear upward trend in win rate with experience. The distribution is moderately tight, supporting the statistical robustness of the learning effect observed in the regression. This confirms that skill acquisition plays a substantial role in Rummy, where experienced players tend to perform better over time—despite the inherent randomness in card draws.
\section{Analyzing Ludo}
Ludo is a multiplayer board game for 2 to 4 players, derived from the Indian game Pachisi. Each player has four tokens that must travel from a starting area to the center of the board by moving around a fixed path. Movement is determined by rolling a six-sided die, and players can capture opponents’ tokens to send them back to the start. While the game is heavily influenced by chance due to dice rolls, strategic choices—such as which token to move and when to block or chase opponents—also play a role. However, the high randomness-to-decision ratio makes it less skill-dominant compared to other games.
\subsection{Ludo Experimental Data}
For Ludo, we have used prelaunch experimental data with a small number of players (see Table \ref{tab:data-table}. The regression results are presented in table \ref{tab:regression-coefficients}. Here previous performance plays an important role, in addition we also see the number of games in the second tranche being a serious factor with a negative sign. This is probably due to fatigue.
The positive contribution of previous experience is also mildly important.

\begin{table}[h!]
\centering
\resizebox{\columnwidth}{!}{%
\begin{tabular}{|c|p{2cm}|p{2cm}|p{2.5cm}|p{2.5cm}|p{2cm}|p{2cm}|}
\hline
\textbf{Player ID} & \textbf{1st Half Win \% ($w_1$)} & \textbf{2nd Half Win \% ($w_2$)} & \textbf{$\phi^{-1}(w_1)$} & \textbf{$\phi^{-1}(w_2)$} & \textbf{1st Half Games ($e_1$)} & \textbf{2nd Half Games ($e_2$)} \\
\hline
2113 & 0.46 & 0.60 & -0.09 & 0.25 & 26 & 25 \\
2708 & 0.35 & 0.25 & -0.38 & -0.67 & 40 & 40 \\
3179 & 0.52 & 0.62 & 0.04 & 0.31 & 68 & 67 \\
3421 & 0.52 & 0.61 & 0.05 & 0.29 & 23 & 22 \\
3544 & 0.31 & 0.37 & -0.51 & -0.32 & 41 & 40 \\
3671 & 0.61 & 0.61 & 0.27 & 0.29 & 70 & 70 \\
3276 & 0.50 & 0.60 & 0.00 & 0.25 & 40 & 40 \\
3296 & 0.52 & 0.53 & 0.06 & 0.08 & 75 & 76 \\
3449 & 0.53 & 0.55 & 0.08 & 0.12 & 72 & 73 \\
3480 & 0.42 & 0.50 & -0.19 & 0.00 & 26 & 26 \\
3581 & 0.27 & 0.27 & -0.60 & -0.60 & 22 & 22 \\
3646 & 0.45 & 0.49 & -0.13 & -0.03 & 41 & 41 \\
\hline
\end{tabular}
}
\caption{Data for Regression: First 6 IDS are from 24 moves, last 6 from 30 moves.}
\label{tab:data-table}
\end{table}

\begin{table}[h!]
\centering
\resizebox{\columnwidth}{!}{%
\begin{tabular}{|c|c|c|c|c|}
\hline
\textbf{Coefficients} & \textbf{Estimate} & \textbf{Std. Error} & \textbf{Test Statistic} & \textbf{p-value} \\
\hline
Intercept & 0.34 & 0.14 & 2.39 & 0.04 \\
$\phi^{-1}$(First Half Win Proportion) & 1.36 & 0.203 & 6.67 & 0.0001 \\
First Half Games & 0.16 & 0.086 & 1.84 & 0.103 \\
Second Half Games & -0.16350 & 0.086 & -1.89 & 0.095 \\
\hline
\end{tabular}
}
\caption{Linear Regression Summary:Multiple R$^{2}$ = 0.878}
\label{tab:regression-coefficients}
\end{table}

\subsection{Ludo Online Data}
\label{regression}
This study uses data obtained from a new variant of the Ludo game as an integrated offering on the My11Circle (MEC, \url{https://www.my11circle.com/}), a fantasy gaming platform. The primary objective of launching the Ludo game on the MEC platform has been to collect the real user game play data. We use this data to conduct our Study. The dataset consists of a total of 265062 players, who have played a total of 3033645 games among themselves during $5^{th}$ June to $23^{rd}$ July. 
The subsetted dataset used for the actual analysis includes players who have played at least 10 games in both months, consists of a total of 16197 players, who have played a total of 2018935 games, and  1191656 games among themselves during $5^{th}$ June to $23^{rd}$ July.

\begin{table}[h!]
\centering
\resizebox{\columnwidth}{!}{%
\begin{tabular}{|c|c|c|c|c|}
  \hline
Variable & Estimate & Std.Error & t-statistic & p-value \\ 
  \hline
Intercept & -0.0988 & 0.0028 & -35.038 & 0 \\ 
$\phi^{-1}$(First Half Win \%)  & 0.2823 & 0.0081 & 34.866 & 0 \\ 
First Half Experience & 0.007 & 0.0048 & 1.459 & 0.143 \\ 
Second Half Experience & 0.0899 & 0.0047 & 18.928 & 0 \\ 
   \hline
\end{tabular}
}
\caption{Regression Output: Multiple  $R^2:  0.1364$}
\label{tab:ludo_regression}
\end{table}

The regression results in table \ref{tab:ludo_regression} firmly show that the role of previous experience (learning) and previous performance (innate skill) is quite prominent in determining current performance of a player. This is a strong indicator of the importance of skill in this Ludo game.

\subsection{Bootstrap Results}
\begin{figure}[h!]
    \centering
    \includegraphics[width= 0.40\textwidth]{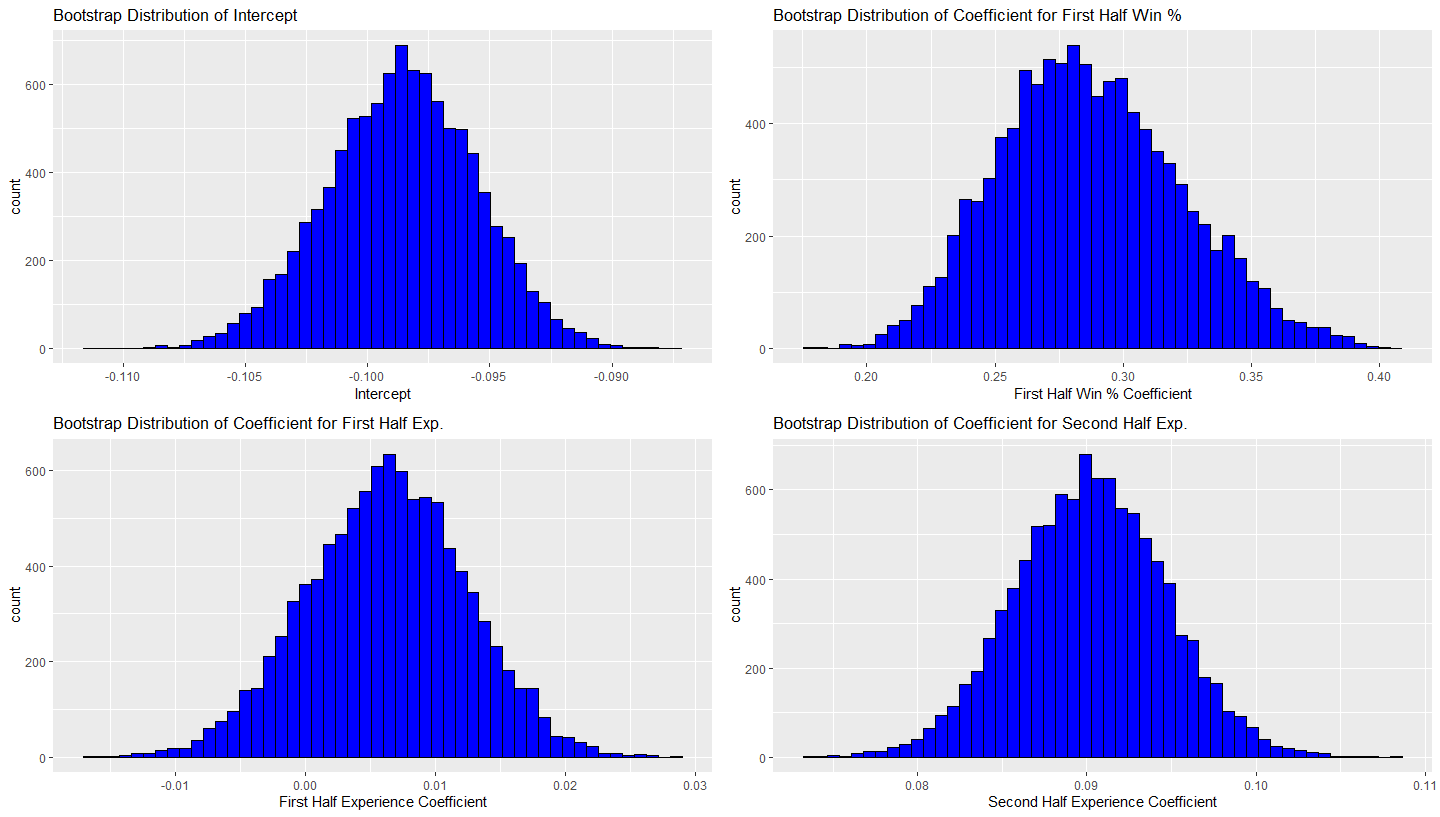} 
    \caption{Histogram of Bootstrap Estimates of Regression Coefficients}
    \label{fig:ludo_bootstrap}
\end{figure}

To understand the robustness of our results, we now perform a 10000 fold bootstrap analysis of the regressions results. The consistency of the results through the bootstrap analysis, as evidenced in the bounds of the coefficients in table \ref{tab:Ludo_online_data} and tightness of the histograms in figure \ref{fig:ludo_bootstrap},  further strengthen our conclusions. 

\begin{table}[h!]
\centering
\resizebox{\columnwidth}{!}{%
\begin{tabular}{|c|c|c|c|c|}
\hline
Variable & Estimate & Bootstrap Mean & Bootstrap Variance & Bootstrap C.I\\
\hline
Intercept & -0.0988 & -0.0985 & 0.00009 & (-0.0991,-0.0982)\\
$\phi^{-1}$(First Half Win \%) & 0.2823  & 0.2861 & 0.0013 & (0.2819,0.2908)\\
First Half Experience & 0.007 & 0.0062 & 0.00003 & (0.0056,0.0072)\\
Second Half Experience & 0.0899  & 0.0902 & 0.0001 & (0.0893,0.0907)\\
\hline
\end{tabular}
}
\caption{Bootstrap regression summary for 10000 bootstrap estimates}
\label{tab:Ludo_online_data}
\end{table}

\FloatBarrier
In the case of Ludo, the bootstrap slope distribution is centered around a slightly positive mean but with a relatively wide spread. This indicates that while some users may improve with experience, the association between experience and win rate is weak and highly variable. These findings are in line with the regression results and reflect the game's high dependence on dice-based randomness, limiting the influence of skill over repeated play.
\section{Analyzing Teen Patti}
Teen Patti is a gambling-style card game popular in South Asia, often compared to three-card poker. Typically played by 3 to 6 players, each participant is dealt three cards face down. Players bet in rounds, deciding whether to fold, call, or raise based on their hand strength or strategic bluffing. The game has hidden information, high variance, and limited opportunities for strategic depth within a single round. While repeated play may reward probabilistic thinking and bluffing skill, short-term outcomes are heavily driven by chance, especially due to the lack of visibility of other players’ cards.

We consider 5 player games only, as these are the most common. In both the versions, previous performance turns out to be an important factor which probably is also related to innate skills. The learning feature is supported by  data as previous experience has a negative impact on current performance.
We analyze the two most popular variations of this game, No Limit and Regular. 
\begin{table}[h!]
\centering
\resizebox{\columnwidth}{!}{%
\begin{tabular}{|c|c|c|c|c|}
\hline
\textbf{Variable} & \textbf{Estimate} & \textbf{Std. Error} & \textbf{Test Statistic} & \textbf{p-value} \\ \hline
Intercept & -0.839 & 0.009 & -89.827 & 0 \\ 
$\phi^{-1}$(First Half Win \%) & 0.357 & 0.004 & 81.99 & 0\\ 
First Half Experience & -5.716 & 0.589 & -9.693 & 0 \\ 
Second Half Experience & 5.9159 & 0.595 & 9.936 & 0\\ 
\hline
\end{tabular}
}
\caption{Regression Summary for \textbf{No Limit} games: multiple $R^{2}$ = 0.116 based on 53621 users}
\label{tab:TeenPatti_nolimit}
\end{table}

\begin{table}[h!]
\centering
\resizebox{\columnwidth}{!}{%
\begin{tabular}{|c|c|c|c|c|}
\hline
\textbf{Variable}  & \textbf{Estimate} & \textbf{Std. Error} & \textbf{t-statistic} & \textbf{p-value} \\   \hline
Intercept & -0.922 & 0.0103 & -89.515 & 0 \\ 
$\phi^{-1}$(First Half Win \%) & 0.226 & 0.006 & 38.077 & 0\\ 
First Half Experience & -0.886 & 0.723 & -1.225 & 0.221 \\ 
Second Half Experience & 1.053 & 0.722 & 1.46 & 0.144 \\ 
\hline
\end{tabular}
}
\caption{Regression Summary for \textbf{Regular} Games: Multiple $R^{2}$ = 0.038 based on 38927 users}
\label{tab:TeenPatti_regular}
\end{table}
Table \ref{tab:TeenPatti_nolimit}) presents the regression summary for the "No Limit" version of Teen Patti, based on a large dataset of over 53,000 users. The model indicates that the first-half win proportion has a strong and highly significant positive effect on current performance, suggesting that initial success—potentially reflective of innate skill or favorable card sequences—plays an important role. Interestingly, the coefficients for first-half experience and second-half experience are significant but of opposite signs: the negative coefficient for first-half experience may suggest overconfidence or fatigue effects, while the positive second-half experience reflects adaptive learning during the session. The model's $R^2=0.116$ shows modest explanatory power, consistent with Teen Patti’s stochastic nature.

Table \ref{tab:TeenPatti_regular} summarizes the regression results for the "Regular" version of Teen Patti, involving nearly 39,000 users. Here, the first-half win proportion remains a statistically significant positive predictor, reinforcing the importance of initial performance. However, experience-based variables (both first and second half) are statistically insignificant, suggesting little evidence of strategic learning or sustained improvement. This aligns with the perception that Teen Patti outcomes are primarily governed by chance, especially when gameplay is not iterative or memory-based. The low $R^2=0.038$ further supports this interpretation, highlighting the difficulty in predicting success from past behavior in this version of the game.

\subsection{Bootstrap Results}
\begin{figure}[h!]
    \centering
    \includegraphics[width= 0.5\textwidth]{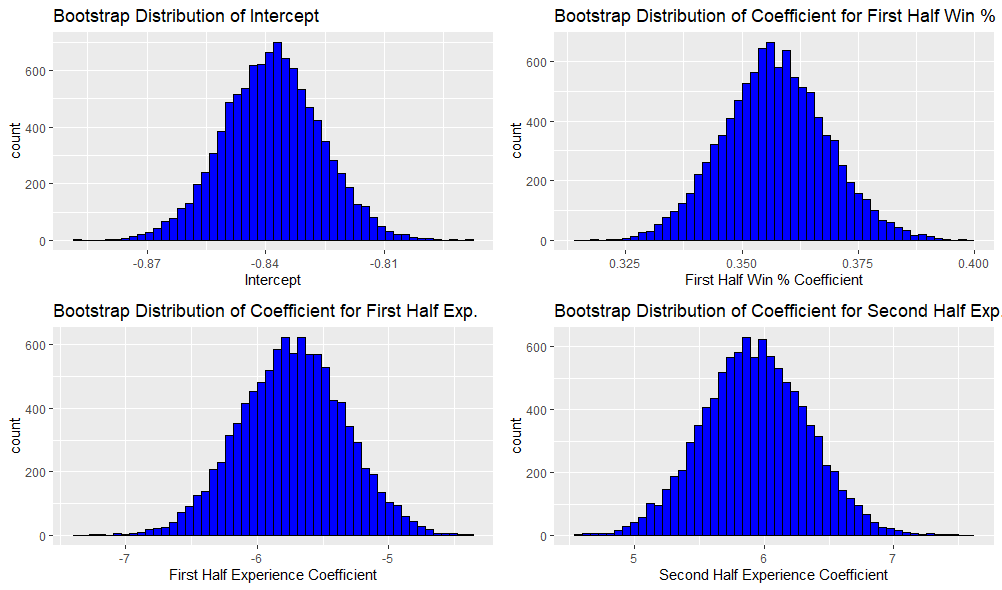} 
    \caption{Histogram of Bootstrap Estimates of Regression Coefficients (Teen Patti-No Limit)}
    \label{fig:TP_NL_Bootstrap}
\end{figure}

First we consider the {\bf No Limit} case.
Bootstrap results (see range of estimates in table \ref{tab:NoLimit_online_data} and histograms in figure \ref{fig:TP_NL_Bootstrap})
confirm the robustness of the regression analysis, indicating that the regression coefficients are likely within the estimated range. The bootstrap estimates exhibit minimal bias, further supporting the reliability of the findings.

\begin{table}[h!]
\centering
\resizebox{\columnwidth}{!}{%
\begin{tabular}{|c|c|c|c|c|}
\hline
Variable & Estimate & Bootstrap Mean & Bootstrap Variance & Bootstrap C.I\\
\hline
Intercept & -0.839 & -0.838 & 0.0001 & (-0.863,-0.814)\\
$\phi^{-1}$(First Half Win \%) & 0.357  & 0.357 & 0.0001 & (0.335,0.379)\\
First Half Experience & -5.716 & -5.722 & 0.159 & (-6.517,-4.949)\\
Second Half Experience & 5.915  & 5.922 & 0.164 & (5.142,6.728)\\
\hline
\end{tabular}
}
\caption{Bootstrap regression summary for 10000 bootstrap estimates (Teen Patti- No Limit)}
\label{tab:NoLimit_online_data}
\end{table}
\FloatBarrier
\FloatBarrier
The bootstrap distribution of slope estimates for Teen Patti (No Limit) exhibits a moderate positive skew, indicating some evidence of performance improvement with experience. However, the spread of the distribution is relatively wide, reflecting variability in learning patterns across users. This suggests that while experience can enhance performance in some cases, outcomes remain partially influenced by chance. The results are consistent with the regression findings, highlighting a modest skill component in the No Limit version of Teen Patti.

Next we analyze the {\bf Regular} one.

\begin{table}[h!]
\centering
\resizebox{\columnwidth}{!}{%
\begin{tabular}{|c|c|c|c|c|}
\hline
Variable & Estimate & Bootstrap Mean & Bootstrap Variance & Bootstrap C.I\\
\hline
Intercept & -0.922 & -0.922 & 0.0002 & (-0.952, -0.891)\\
$\phi^{-1}$(First Half Win \%) & 0.226  & 0.226 & 0.0002 & (0.198, 0.255)\\
First Half Experience & -0.886 & -0.898 & 0.544 & (-2.196,0.505)\\
Second Half Experience & 1.053 & 1.067 & 0.565 & (-0.235,2.176)\\
\hline
\end{tabular}
}
\caption{Bootstrap regression summary for 10000 bootstrap estimates (Teen Patti-Regular)}
\label{tab:Regular_online_data}
\end{table}
\FloatBarrier
\begin{figure}[h!]
    \centering
    \includegraphics[width= 0.45\textwidth]{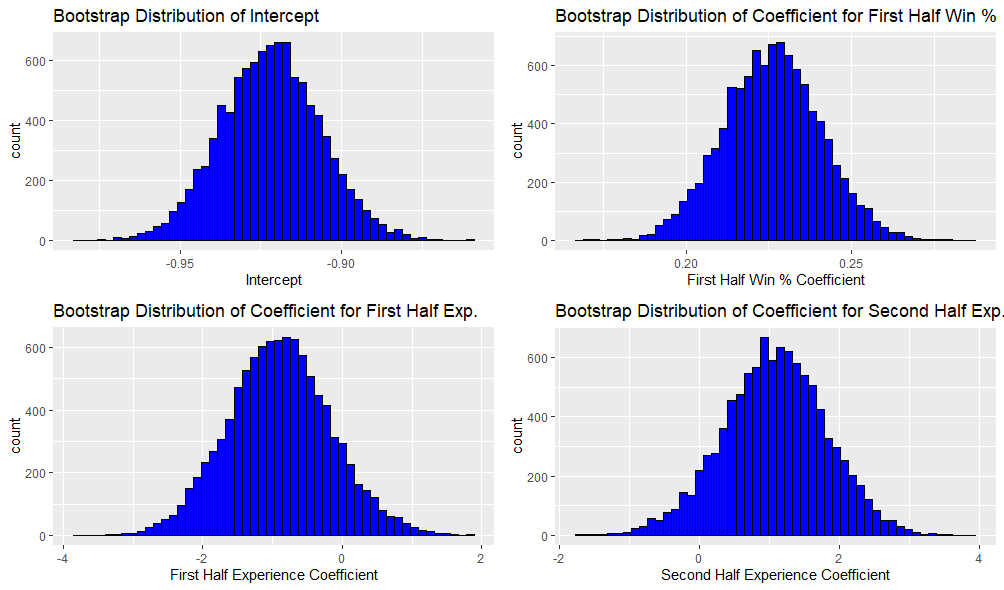} 
    \caption{Histogram of Bootstrap Estimates of Regression Coefficients (Teen Patti-Regular)}
    \label{fig:TP_Reg_Bootstrap}
\end{figure}
\FloatBarrier
The bootstrap slopes for Teen Patti (Regular) (Table\ref{tab:Regular_online_data}, Figure \ref{fig:TP_Reg_Bootstrap}) are centered around zero with a wide spread, indicating weak or no consistent relationship between experience and performance. This supports the interpretation that outcomes in this version are largely driven by chance.
\section{Methodology to Generate the Final Skill Score}
\label{Overall}
\subsection{Qualitative Considerations}
In this report, our main objective is to draw comparative statements on the effect of luck versus skill on the popular card and board games considered (Tables \ref{tab:Chess_rapid2} to \ref{tab:TeenPatti_regular}). The initial discussion revolves around the importance of ''innate skill" (performance in general) and ''learning" (experience: learning by doing). Looking at the comparative results in Table \ref{tab:comparative} we can assess this for the games under consideration. We certainly look at \textbf{Chess} as the benchmark case. Here rating proxies for skill acquired through experience. The regression results are very clear about the role of previous performance and rating. 

Among the other three games, first consider \textbf{Teen Patti}, as this is popularly considered to be the ``least skilled'' game.  Here we see a significant role of ``innate skill" but no learning effect. This skill may be attitudinal (bluffing) or cognitive (recognising the behavioural markers of the opponent). But, there is no strategic learning. 

Moving on to \textbf{Rummy} the results are very different. Again the relevant ''innate skill" here is the choice of strategy in selection and retention of cards, and there is a significant learning effect. In fact this is true even for current experience (albeit at a lower level of significance). Contrasted with this, if we consider \textbf{Ludo}, we again see both skill as well as learning effect but the effect is now slightly weaker. In fact, the experience effect in Ludo is somewhat mixed with a positive and negative sign.

\begin{table}[h!]
    \centering
    \resizebox{\columnwidth}{!}{%
    \begin{tabular}{|l|>{\centering\arraybackslash}p{2.2cm}|>{\centering\arraybackslash}p{2.2cm}|>{\centering\arraybackslash}p{2.2cm}|c|}
    \hline
         Game&  Previous Performance&  Previous Experience&  Current Experience& $R^2$\\
         \hline
         {\bf Ludo} &&&&\\
         Experiment&  +&  + (*)&  - (*)& 0.88\\
         Online & + & + (*) & + & 0.14 \\
         \hline
         {\bf Chess} (Real)&  &  &  & \\
         Classical&  &  &  & 0.48\\
         Rapid (1y)&  -&  +&  + (Rating)& 0.27\\
         Rapid (2y)&  + (*)&  &  + (Rating)& 0.61\\
         Chess (Online) & + & +& +& 0.57 \\
         \hline
         {\bf Rummy} (online)&  +&  +&  + (*)& 0.013\\
         \hline
         {\bf Teen Patti} (online)&  &  &  & \\
         No Limit&  +&  -&  +& 0.15\\
         Regular&+&&& 0.05\\
         \hline
    \end{tabular}
    }
    \caption{Overall Comparison (Significance at 5\% level or *:$|t| > 1$)}
    \label{tab:comparative}
\end{table}

We see that performance in all the games under consideration rely on innate skill (player heterogeneity), which is one indicator of skill base. 
Interim findings also suggest that an one shot game like {\bf Teen Patti} is less amenable to learning compared to the repeated interaction games, e.g. {\bf Rummy} and the two board games considered. Among the board games, {\bf Ludo} illustrates a slightly weaker effect of learning compared to {\bf Rummy} based on the importance of previous experience in terms of the t-stat. See tables \ref{tab:Rummy} and \ref{tab:regression-coefficients}, although the difference is very minor.
Finally, of course, {\bf Chess} is the most clear cut skill based game. Based on our limited analysis, we can tentatively rank the four games in order of increasing skill and learning dependence as
\begin{center}
     \textbf{Teen Patti} $\prec$ \textbf{Ludo} $\prec$ \textbf{Rummy} $\prec$ \textbf{Chess}.
\end{center}
However, as we have noted earlier, \textbf{Ludo} and \textbf{Rummy} show very close quantitative scores in terms of our analysis.

\subsection{The Skill Score}
We now attempt a scoring exercise on a $[0, 1]$ scale for ``Skill''. This is tentative as it requires several subjective choices.

Again, we use the regressions results from tables \ref{tab:regression-coefficients} for Ludo, \ref{tab:Chess_rapid2} for Chess in the Rapid format with 2 years data, \ref{tab:Rummy} for Rummy and \ref{tab:TeenPatti_nolimit} for Teen Patti of the ``No Limit'' variety to represent the four games under consideration.

To avoid scaling issues, we use the  t-statistics for the relevant factors instead of the regression coefficients as constituents of the score. Let 

$t_1$ denote the t-statistics for performance in first half 

$t_2$ denote the t-statistics for no. of games in first half 

$t_3$ denote the t-statistics for no. of games in second half \\

Then, using the properties of the Normal distribution as we are working with fairly large samples, we define the normalised (in [0, 1]) versions as

\begin{equation}
\label{scalingeqn}
x_i = \left\{ \begin{array}{lll}
0 & if & t_i < 0 \\
t_i/4 & if & 0 \leq t_i \leq 2 \\
(t_i+1)/6 & if & 2 < t_i \leq 5 \\
1 & if & 5 < t_i
\end{array} , i = 1, 2, 3. \right.
\end{equation}

The choice of the cut-offs, 2 and 5, are based on the level of significance (corresponding to $10^{-2}$ and $10^{-6}$ respectively). These are subjective choices and can be reconsidered. The choice of slope ($1/4$ and $1/6$ is also subjective, the only objective decision being to award a higher importance to the initial phase (a concave evaluation).

Finally, to define a composite skill score, we propose assigning a weight of 2:1 to the effect of past experience (learning) relative to innate skill. This choice reflects our belief that a player’s ability to improve with repeated play is a more compelling indicator of skill than a high initial win rate alone, which could be due to chance or prior unseen experience. Emphasizing learning aligns with the broader notion of skill as something that develops through mastery, adaptation, and strategic refinement. Moreover, in legal and policy contexts, games are more likely to be classified as skill-based when success can be consistently linked to effort and experience rather than initial performance. In turn, innate skill in the form of past performance should also receive twice as much weight as current experience, as current experience may be confounding learning and fatigue. Thus, we define 
the skill score as 
\begin{equation}
\label{wtgeqn}
    s = \frac{2x_1 + 4x_2 + x_3}{7} \in [0, 1].
\end{equation}

While we adopt this weighting in this study, the framework is flexible and can accommodate alternative preferences depending on the context e.g., game design analysis, player behavior modeling, or regulatory thresholds. We hope that we can objectify this choice with a more comprehensive analysis in future work.

Below we illustrate with the games considered here in table \ref{score}. The Scores obtained almost confirms our ranking posited above, the only exception being Ludo scores slightly higher than Rummy now. As these two numbers are again very close to each other, we think that the results may be sensitive to the choice of cut-offs and weightages in the score formula. 

\begin{table}[h!]
    \centering
    \resizebox{\columnwidth}{!}{%
    \begin{tabular}{|c|c|c|c|c|c|c|c|}
    \hline
           \multicolumn{2}{|c}{Ludo}&  \multicolumn{2}{|c}{Chess}&  \multicolumn{2}{|c}{Rummy}&  \multicolumn{2}{|c|}{Teen Patti}\\
         \hline
           t-stat ($t$)&  scaled ($x$) &  t-stat ($t$)&  scaled ($x$) &  t-stat ($t$)&  scaled ($x$) &  t-stat ($t$) & scaled ($x$)
\\
\hline
           34.866&  1.00&  9.12&  1.00&  15.738&  1.00&  93.692& 1.00
\\
           1.459&  0.37&  20.23&  1.00&  2.149&  0.52&  -9& 0.00
\\
           18.928&  1.00&  11.75&  1.00&  1.065&  0.27&  9& 1.00
\\
\hline
           &  $s =$ \textbf{0.64}&  & $s =$ \textbf{1.00}&  & $s =$ \textbf{0.62}&  & $s =$ \textbf{0.43}
\\
\hline
    \end{tabular}
    }
    \caption{Skill Score for Games considered}
    \label{score}
\end{table}

\subsection{Resampling Results on Skill Score}
Based on the findings above, we have reason to investigate the choice of score function further. In this part of the report we consider various choices for the cut-off and weightage to make the functional form of the skill score flexible subject to the intuitive restrictions. To illustrate, we use a range of weightages (e.g. between 0.5 and 0.6 in place of $4/7$). \\
One may also use different non-linear transformation $x_i (t_i) : R \rightarrow [0,1]$ on the t-stat. As any smooth transformation on a suitable domain can be well approximated by piece-wise linear functions, we claim that by varying the cut-off and weightages, the proposed score function provides sufficient variety.

The results are  not only dependent on the choice of the score function. Sampling fluctuation inherent in the data may also influence our scores and hence ranking. So we also carry out a bootstrap analysis with our regression model. We will get a range of t-stat based on the range of regression coefficients from bootstrap analysis. This will give us a range of $t_i$ values and hence a range of $x_i$ values. Thus, instead of a point estimate for the score function $s$ for each game, it will have a range of values. We can then compare distribution of the score values to arrive at a comprehensive ranking template. The details are available in the following table.

\begin{table}[h!]
    \centering
     \resizebox{\columnwidth}{!}{%
    \begin{tabular}{|c|c|}
    \hline
        Choice of cut-offs &Transformation $x_i (t_i) : R \rightarrow [0,1]$   \\ \hline
       2,5  &  $x_i = \left\{ \begin{array}{lll}
0 & if & t_i < 0 \\
t_i/4 & if & 0 \leq t_i \leq 2 \\
(t_i+1)/6 & if & 2 < t_i \leq 5 \\
1 & if & 5 < t_i
\end{array},  i = 1, 2, 3. \right. $\\ \hline
       1.5, 4.5 & $x_i = \left\{ \begin{array}{lll}
0 & if & t_i < 0 \\
t_i/3 & if & 0 \leq t_i \leq 1.5 \\
(t_i+1.5)/6 & if & 1.5 < t_i \leq 4.5 \\
1 & if & 4.5 < t_i
\end{array},  i = 1, 2, 3. \right. $\\ \hline
       1.5, 5.5 & $x_i = \left\{ \begin{array}{lll}
0 & if & t_i < 0 \\
t_i/3 & if & 0 \leq t_i \leq 1.5 \\
(t_i+2.5)/8 & if & 1.5 < t_i \leq 5.5 \\
1 & if & 5.5 < t_i
\end{array},  i = 1, 2, 3. \right. $\\ \hline
       2.5, 4.5 & $x_i = \left\{ \begin{array}{lll}
0 & if & t_i < 0 \\
t_i/5 & if & 0 \leq t_i \leq 2.5 \\
(t_i-0.5)/4 & if & 2.5 < t_i \leq 4.5 \\
1 & if & 4.5 < t_i
\end{array},  i = 1, 2, 3. \right. $\\ \hline
       2.5, 5.5 & $x_i = \left\{ \begin{array}{lll}
0 & if & t_i < 0 \\
t_i/5 & if & 0 \leq t_i \leq 2.5 \\
(t_i+0.5)/6 & if & 2.5 < t_i \leq 5.5 \\
1 & if & 5.5 < t_i
\end{array},  i = 1, 2, 3. \right. $\\ \hline
 $(a,b)$, $a \in (1.5,2.5)$ \& $b \in (4.5,5.5)$& $x_i = \left\{ \begin{array}{lll}
0 & if & t_i < 0 \\
t_i/2a & if & 0 \leq t_i \leq a \\
\frac{t_i+(b-2a)}{2(b-a)} & if & a < t_i \leq b \\
1 & if & b < t_i
\end{array},  i = 1, 2, 3. \right. $\\ \hline
    \end{tabular}
    }
    \caption{Possible Transformations $x_i (t_i) : R \rightarrow [0,1]$}
    \label{tab:Transformations}
\end{table}

Note that the transformations posited in table \ref{tab:Transformations} covers both convex and concave formulations. It is convex (concave) if $b > (<) 2a.$ Thus, when we generalise the discussion to smooth functions, we can cover both the aspects. Further, we also allow a range of weightages in the table \ref{tab: Range of Weightages} which can approximate the domain under our intuitive restrictions quite well.

\begin{table}[h!]
    \centering
    \begin{tabular}{|c|c|c|}
    \hline
       $\mathbf{x_1}$ & $\mathbf{x_2}$ & $\mathbf{x_3}$  \\ \hline
        $\frac{2}{7}$ & $\frac{4}{7}$ & $\frac{1}{7}$
        \\ \hline
       0.25&0.5 &0.25 \\ \hline 
     0.3 &0.6 & 0.1 \\ \hline 
  $\frac{a}{2}$   & $a$,   $a \in (0.5,0.6)$ & $1 - \frac{3a}{2}$ \\ \hline
    \end{tabular}
    \caption{Possible Range of Weightages}
    \label{tab: Range of Weightages}
\end{table}

\begin{figure*}[ht!]
    \centering
\includegraphics[width=1\textwidth]{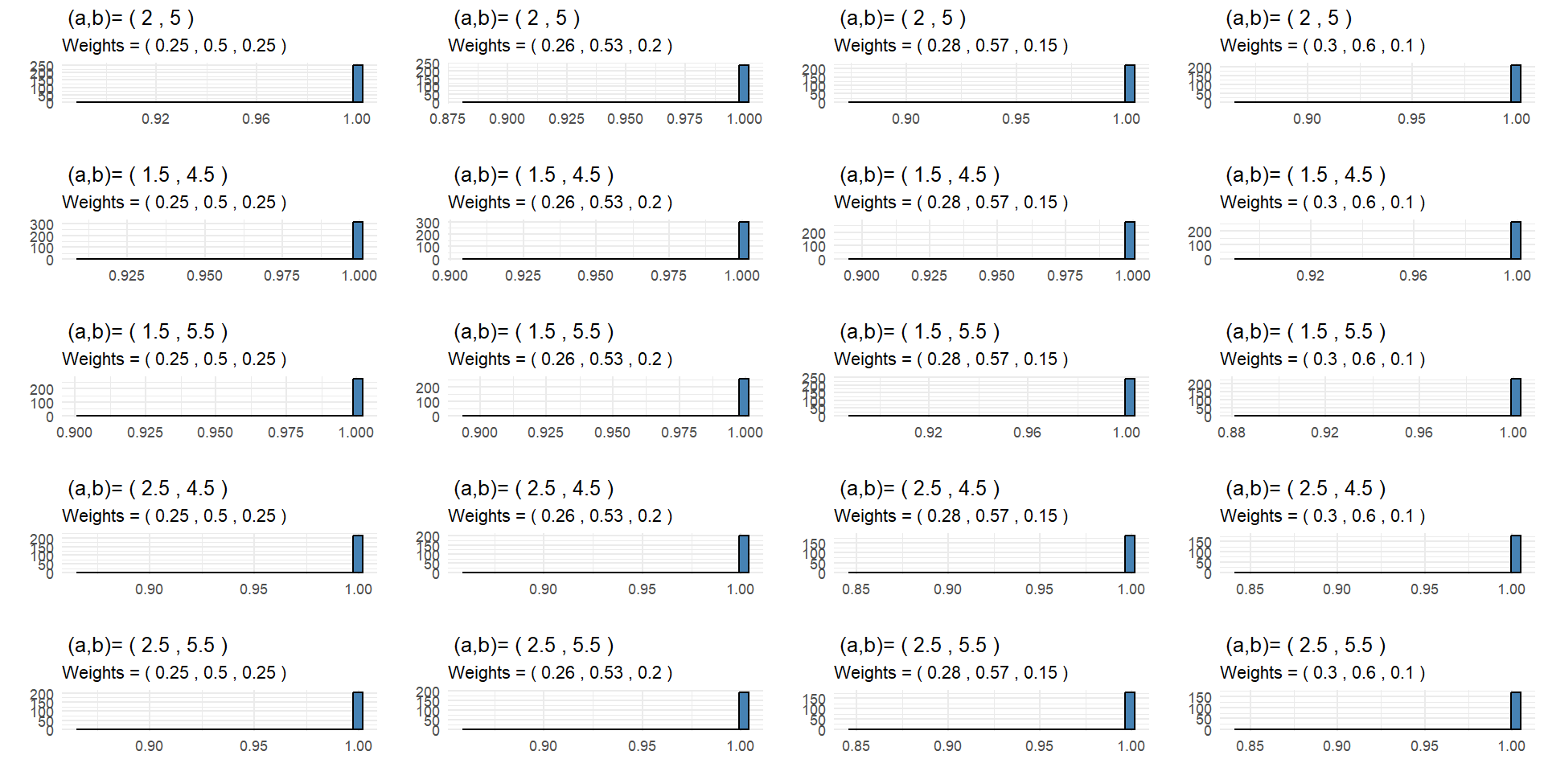}
    \caption{Distribution of Scores for Various Transformations for Chess Bootstrapped Data}
    \label{fig:Chess_Scores_Transformations}
\end{figure*}

\begin{figure*}[ht!]
    \centering
\includegraphics[width=\textwidth]{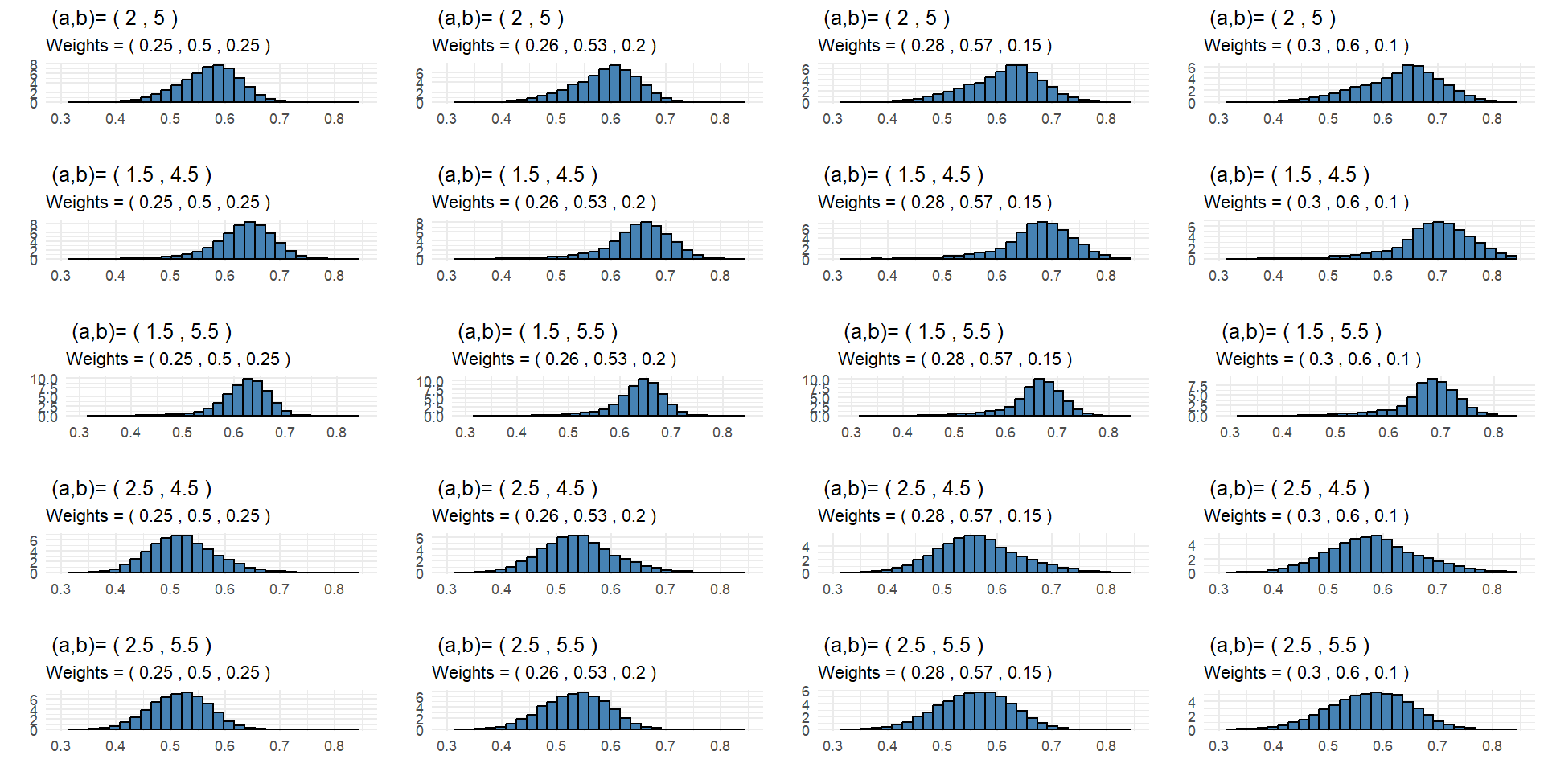}
    \caption{Distribution of Scores for Various Transformations for 2-Player Rummy Bootstrapped Data}
    \label{fig:Rummy2_Scores_Transformations}
\end{figure*}

\begin{figure*}[ht!]
    \centering
\includegraphics[width=\textwidth]{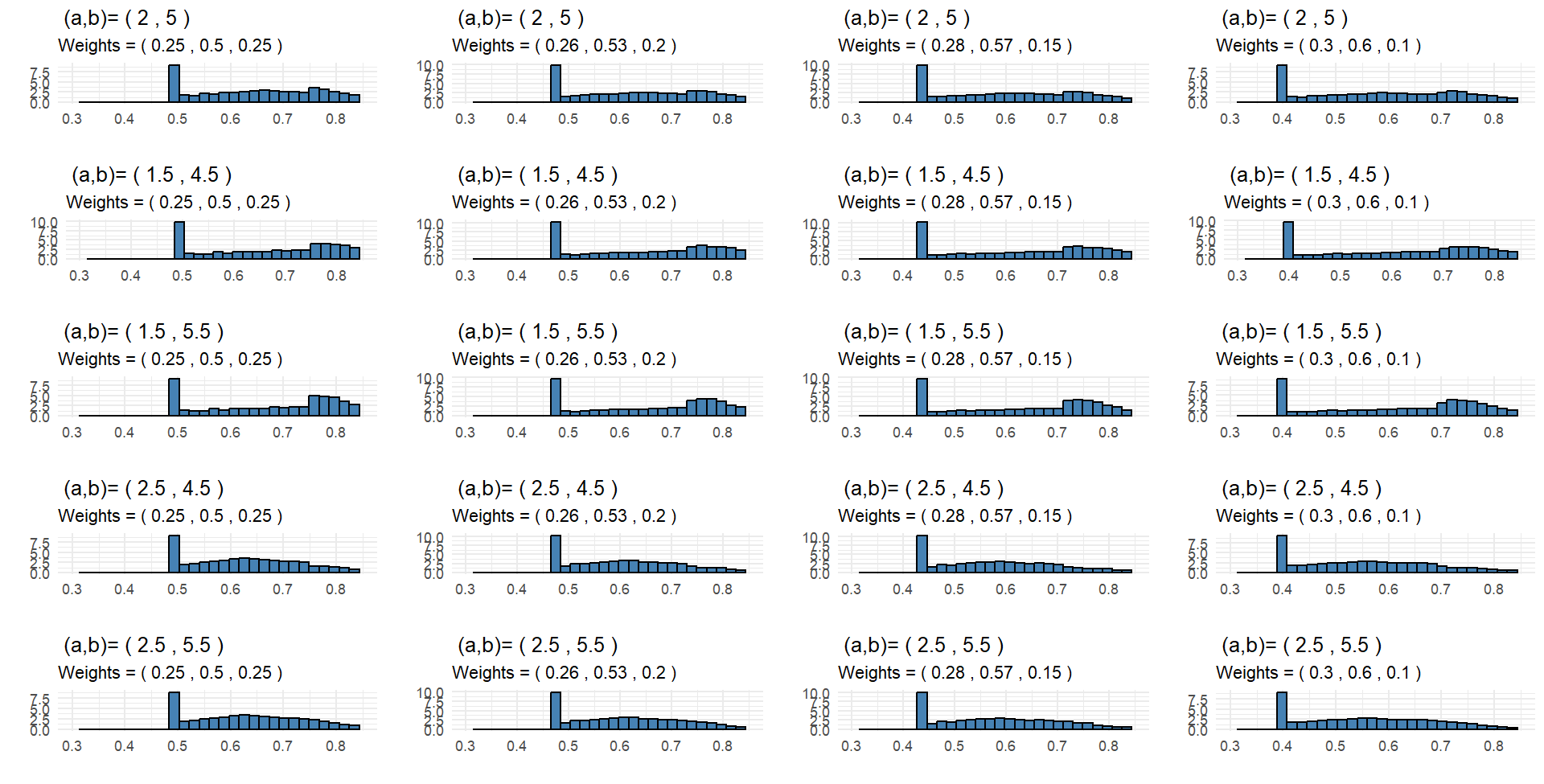}
    \caption{Distribution of Scores for Various Transformations for Ludo Bootstrapped Data}
    \label{fig:Ludo_Scores_Transformations}
\end{figure*}

\begin{figure}[ht!]
    \centering
\includegraphics[width=\textwidth]{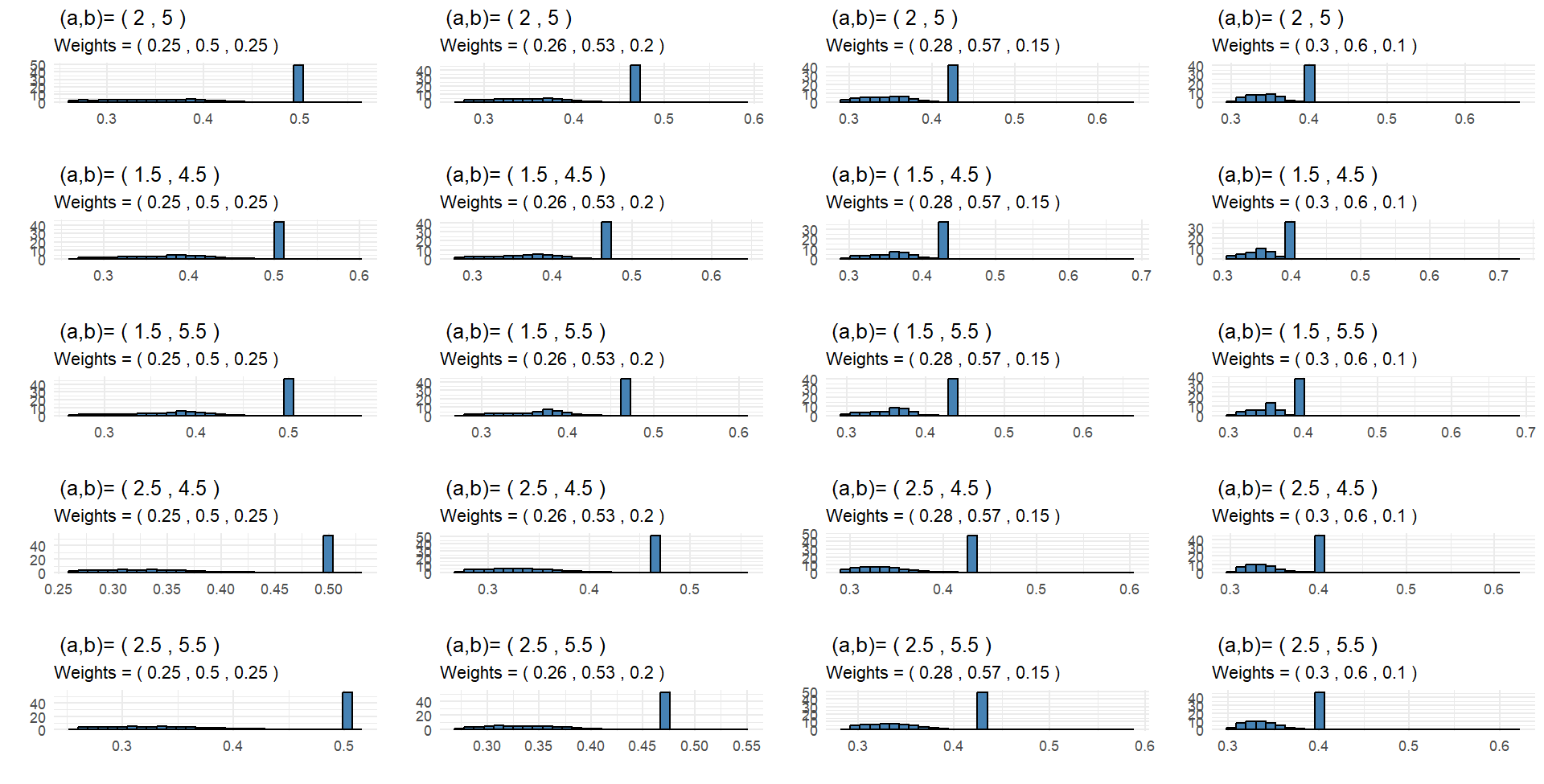}
\caption{Distribution of Scores for Various Transformations for Teen Patti Bootstrapped Data}  
\label{fig:TeenPatti_Scores_Transformations}
\end{figure}

For each of the games we use 10000 bootstrap samples to generate the t-statistics. Hence, we get a distribution of scores for each of the combinations in table \ref{tab:Transformations}. These are presented in figures \ref{fig:Chess_Scores_Transformations} for Chess, 
\ref{fig:Rummy2_Scores_Transformations} for Rummy, \ref{fig:Ludo_Scores_Transformations} for Ludo and finally 
\ref{fig:TeenPatti_Scores_Transformations} for Teen Patti at the end of this paper.

As expected, we now see a clearly articulated distribution of score for each game which shifts in centre and spread with the cut-offs and transformations chosen. Except for Chess, which is degenerate at 1. Now our skill quantification and comparison exercise transforms into a (possibly) partial ordering problem as the distributions for different games may now overlap.

\FloatBarrier
\begin{table*}[th!]
    \centering
    {\small
    \begin{tabular}{|c|c|c|c|c|c|}
    \hline
         Weightage&  (a, b)&  Teen Patti&   Rummy&Ludo& Chess\\ \hline
         &  (2,5)&  (0.3, 0.5, 0.5)&   (0.496, 0.575, 0.636)&(0.500, 0.667, 0.826) & (1,1,1)\\
         &  (1.5,4.5)&  (0.318, 0.5, 0.5)&   (0.567, 0.639, 0.698)&(0.500, 0.722, 0.868)& (1,1,1)\\
         (0.25,0.5,0.25)&  (1.5,5.5)&  (0.318, 0.5, 0.5)&   (0.561, 0.624, 0.671)&(0.500, 0.722, 0.838) & (1,1,1)\\
         &  (2.5,4.5) &  (0.291, 0.5, 0.5)&   (0.447, 0.518, 0.602)&(0.500, 0.633, 0.802) & (1,1,1)\\
         &  (2.5,5.5)&  (0.291, 0.5, 0.5)&   (0.447, 0.518, 0.582)&(0.500, 0.633, 0.784)& (1,1,1) \\ \hline
         &  (2,5)
&  (0.308, 0.47, 0.47)&   (0.505, 0.595, 0.659)&(0.470, 0.647, 0.816)& (1,1,1) \\
         &  (1.5,4.5)
&  (0.323, 0.47, 0.47)&   (0.578, 0.657, 0.718)&(0.470, 0.706, 0.860) & (1,1,1)
\\
         (0.26,0.53,0.2)&  (1.5,5.5)
&  (0.323, 0.47, 0.47)&   (0.576, 0.642, 0.689)&(0.470, 0.706, 0.829) & (1,1,1)\\
         &  (2.5,4.5) 
&  (0.299, 0.47, 0.47)&   (0.457, 0.536, 0.629)&(0.470, 0.611, 0.790) & (1,1,1)
\\
 & (2.5,5.5)& (0.299, 0.47, 0.47)& (0.457, 0.536, 0.606)&(0.470, 0.611, 0.772)  &(1,1,1)
\\ \hline
 & (2,5)
& (0.317, 0.43, 0.43)& (0.517, 0.621, 0.694)&(0.430, 0.620, 0.802)  &(1,1,1)
\\
 & (1.5,4.5)
& (0.328, 0.43, 0.43)& (0.59, 0.68, 0.75)&(0.430, 0.683, 0.849)  &(1,1,1)\\
 (0.28,0.57,0.15)& (1.5,5.5)
& (0.328, 0.43, 0.43)& (0.589, 0.665, 0.683)&(0.430, 0.683, 0.816)  &(1,1,1)\\
 & (2.5,4.5) 
& (0.31, 0.43, 0.43)& (0.471, 0.56, 0.666)& (0.430, 0.582, 0.774) &(1,1,1)
\\
 & (2.5,5.5)
& (0.31, 0.43, 0.43)& (0.471, 0.56, 0.641)& (0.430, 0.582, 0.754)&(1,1,1)
\\ \hline
 & (2,5)
& (0.323, 0.4, 0.4)& (0.525, 0.641, 0.721)&(0.400, 0.600, 0.791)  &(1,1,1)
\\
 & 
(1.5,4.5)& (0.33, 0.4, 0.4)& (0.597, 0.699, 0.776)& (0.400, 0.667, 0.841) &(1,1,1)\\
 (0.3,0.6,0.1)& (1.5,5.5)& (0.33, 0.4, 0.4)& (0.596, 0.715, 0.738)&(0.400, 0.667, 0.806) &(1,1,1)
\\
 & 
(2.5,4.5) 
& (0.318, 0.4, 0.4)& (0.48, 0.579, 0.694)&(0.400, 0.560, 0.762)  &(1,1,1)
\\
 & 
(2.5,5.5)
& (0.318, 0.4, 0.4)& (0.48, 0.579, 0.667)& (0.400, 0.560, 0.741) &(1,1,1)
\\ \hline
    \end{tabular}
    }
    \caption{For the four Games considered: Quantiles ($q_{10}, q_{50}, q_{90})$ of Score for  different weightage and (a, b) combinations} 
    \label{score_comparison}
\end{table*}

To simplify our discussion, we produce the median and an 80\% confidence interval for all the score distributions in table \ref{score_comparison} where the degeneracy of Chess score is clear. It is also clear that Teen Patti is less skilled than either Rummy or Ludo (the confidence intervals are practically separated) who are in turn less skilled than Chess. The comparison between Rummy and Ludo is unclear, as already guessed at earlier. The distributions overlap with Ludo having a little more spread and the centre of the distributions (median) also switching order between Rummy and Ludo for alternative parametric configurations. This clearly indicates that these two games are very similar in total skill content as indicated earlier. Of course, the kind of skill required may be quite diverse.

For consistency of comparison, in the end, we have only used data from online real player games based on a large sample of players and games. The analysis of other data sets, real or experimental, are only shown for illustrative purpose.

\section{Concluding Remarks}
It is important to note that some of our regression models yield relatively low $R^2$ values. However, this should not be viewed as undermining the validity of the models or the significance of the predictors. In large-scale observational datasets, especially those capturing human behavior such as game performance outcomes are often influenced by numerous unobserved or idiosyncratic factors (e.g., mood, device used, time of day), which contribute to the residual variance. A low $R^2$ in such contexts does not imply that the identified predictors are uninformative; rather, it reflects the inherent noise in the data and the partial nature of the model's explanatory scope. In fact, when sample sizes are large and predictors are theoretically grounded, the significance of regression coefficients (e.g., p-values, t-statistics, and bootstrap confidence intervals) often offers a more meaningful interpretation than the overall variance explained. This is especially common in behavioral and social science data, where $R^2$ values as low as 0.01–0.05 are considered acceptable if effect sizes are statistically and practically significant \cite{shmueli_to_explain_or_predict},\cite{jacob_cohen_stat__behaviour}. Our use of resampling-based confidence intervals further reinforces the robustness of our coefficient estimates, even when $R^2$ is modest.

The conclusions from our analysis are quite nuanced with Chess being the most skilled (one extreme) and Teen Patti on the other end (but not an extreme score, around 0.45). While Rummy and Ludo are in between, the ranking between them is ambiguous. Depending on cutoffs and weight-age schemes, the score distribution is always overlapping for these two with a wider range for Ludo than Rummy (possibly due to the larger sample of Rummy data). 

While the final skill score relies on subjective choices of cut-offs and weights, we provide multiple formulations to demonstrate robustness. This flexibility is intentional, as different stakeholders (e.g., legal, gaming design, regulatory) may prioritize components like learning rate or innate performance differently. Thus, the framework is not meant to offer a single "correct" number, but rather a statistically grounded range to inform debate.
Thus, as mentioned in the Introduction, we allow for a difference of opinion regarding the skill comparison of Ludo and Rummy. Their skill profiles are unterminated by the other and hence their ranking will change with a change in scaling and weight choice. We feel that this provides a simple but flexible data-driven comparison tool for skill games in general.

While our current framework is built on win statistics, it can be extended to accommodate other definitions of successful gameplay. For example, in cooperative or resource-based games, success may involve minimizing loss, optimizing resources, or sustaining high performance across rounds. Incorporating additional performance metrics would allow for a richer, game-specific evaluation of skill. This is a promising direction for future work.

\bibliographystyle{IEEEtran}

\newpage

\FloatBarrier
\FloatBarrier
\FloatBarrier
\FloatBarrier
\end{document}